\documentclass[twocolumn,english]{revtex4-1}
\usepackage[T1]{fontenc}
\usepackage[latin9]{inputenc}
\usepackage{color}
\usepackage{amsmath}
\usepackage{amssymb}
\usepackage{graphicx}
\usepackage{esint}

\makeatletter
\renewcommand{\fnum@figure}{\textbf{Figure~\thefigure}}
\makeatother
\usepackage{units}

\usepackage{babel}
\begin{document}

\title{Carrier multiplication in graphene under Landau quantization}

\author{Florian Wendler, Andreas Knorr, and Ermin Malic }

\affiliation{Institute of Theoretical Physics, Nonlinear Optics and Quantum Electronics, Technical University Berlin, Hardenbergstr. 36, 10623 Berlin, Germany}

\email{florian.wendler@tu-berlin.de}

\begin{abstract}
Carrier multiplication is a many-particle process giving rise to the generation of multiple electron-hole pairs. This process holds the potential to increase the power conversion efficiency of photovoltaic devices. In graphene, carrier multiplication has been theoretically predicted and recently experimentally observed. However, due to the absence of a bandgap and competing phonon-induced electron-hole recombination, the extraction of charge carriers remains a substantial challenge. Here we present a new strategy to benefit from the gained charge carriers by introducing a Landau quantization that offers a tunable bandgap. Based on microscopic calculations within the framework of the density matrix formalism, we report a significant carrier multiplication in graphene under Landau quantization. Our calculations reveal a high tunability of the effect via externally accessible pump fluence, temperature, and the strength of the magnetic field. 
\end{abstract}
\maketitle

\textbf{Introduction.} In 1961 Schockley and Queisser predicted a
fundamental limit of approximately $30\%$ for the power conversion efficiency of
single junction solar cells \cite{Shockley1961}. Their calculation is based on a
simple model, in which the excess energy of absorbed photons is assumed to
dissipate as heat. This is a good assumption for most conventional
semiconductors, whereas for low-dimensional nanostructures with strong Coulomb
interaction, the excess energy can be exploited to generate additional
electron-hole pairs. In such structures with highly efficient Auger processes,
the Schockley-Queisser limit can be exceeded up to $60\%$ \cite{Landsberg1993}.
Furthermore, the gained multiple electron-hole pairs can also increase the
sensitivity of photodetectors and other optoelectronic devices.

\begin{figure}[!b]
\begin{centering}
\includegraphics[width=85mm]{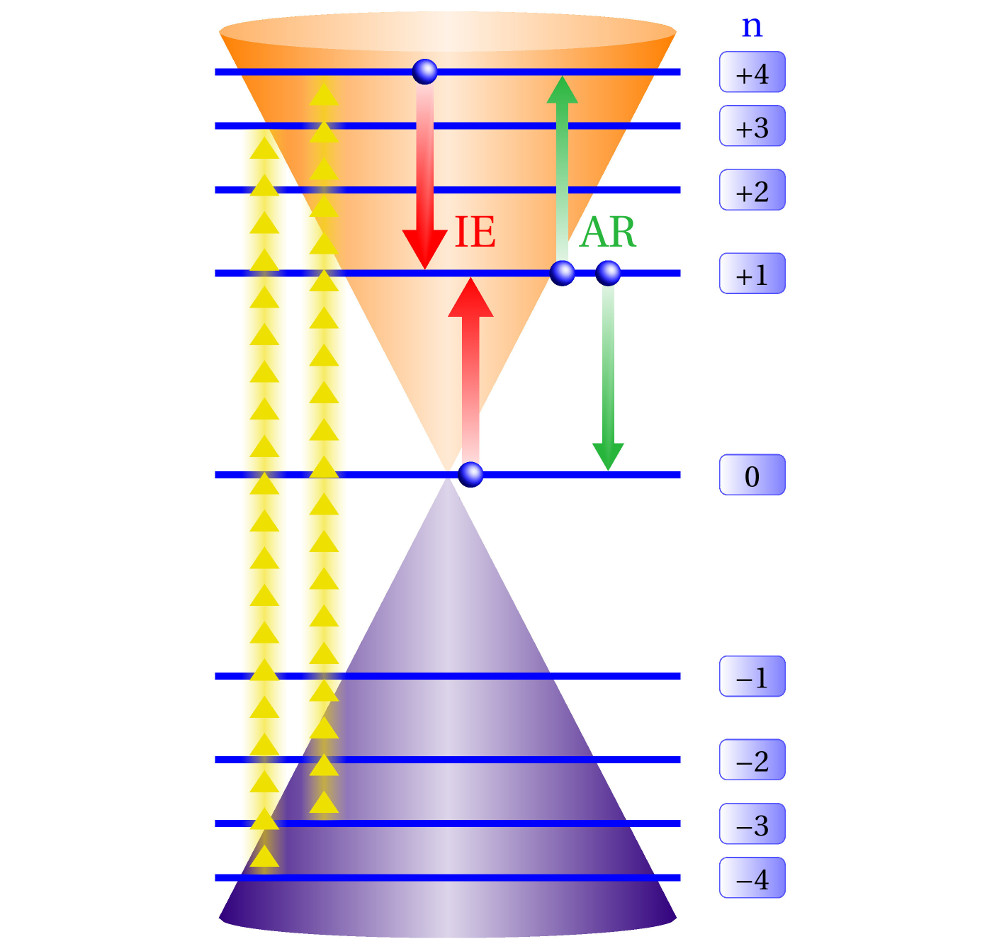} 
\par\end{centering}

\caption{\textbf{Auger scattering in Landau-quantized graphene. }Sketch of the
nine energetically lowest Landau levels of graphene with the Dirac cone in the
background. The optical excitation via a linearly polarized laser pulse induces
the transitions $-4\,\rightarrow\,+3$ and $-3\,\rightarrow\,+4$, cf. the yellow
arrows. The energy-conserving process of impact excitation (IE) involves the
inter-Landau level transition $+4\,\rightarrow\,+1$, which provides the energy
to excite an electron from $n=0$ to $n=+1$ resulting in an increase of the
number of charge carriers, cf. the red arrows. The inverse process of Auger
recombination (AR) is shown by the green arrows. }

\label{fig:sketch} 
\end{figure}

Carrier multiplication (CM) was first theoretically predicted \cite{Nozik2002} and measured \cite{Schaller2004} for semiconductor quantum dots exhibiting a strong spatial confinement in all three directions. Multiplication of photo-excited carriers has also been found in one-dimensional nanostructures, such as carbon nanotubes \cite{Baer2010,Gabor2009}. In graphene, the carrier confinement combined with the linear bandstructure was predicted to account for highly efficient Auger processes \cite{Rana2007} giving rise to a considerable carrier multiplication \cite{Winzer2010_Multiplication,Winzer2012}. Just recently, these predictions were confirmed by Brida et al. within a joint experiment-theory study also including new insights into collinear scattering in graphene \cite{Brida2013}, which has been controversially discussed in literature. However, to realize graphene-based photovoltaic devices, the key problem of charge carrier extraction in gapless nanostructures needs to be solved. Here we suggest a 
strategy based on Landau quantization of graphene to address this substantial challenge, where a magnetic field is used to induce gaps between discrete Landau levels.  Graphene in high magnetic fields has already attracted an immense interest in recent years \cite{Novoselov2005,Zhang2005,Du2009,Goerbig2011Review}. Unlike an ordinary two-dimensional electron gas, it exhibits an unique Landau level structure, where the energy is proportional to the square root of the Landau level index $n=0,\pm1,\pm2,...$ ($E_{n}\propto\pm\sqrt{\left|n\right|}$), cf. Fig. \ref{fig:sketch}. This is a direct consequence of the linear density of states in the low-energy region and results in an anomalous quantum Hall effect \cite{Novoselov2005,Zhang2005}. 

In this Article, we predict the occurrence of carrier multiplication in Landau-quantized graphene, where the energy gap between the Landau levels is of advantage for the extraction of gained charge carriers. The magnetic field can be tuned so that the inter-Landau level spacings do not fit the energy of the most relevant optical phonons resulting in a strongly suppressed carrier-phonon scattering. Although at first sight the non-equidistant Landau level spectrum seems to suppress Auger processes, a number of energetically degenerate inter-Landau level transitions exists providing the basis for efficient Auger scattering, cf. Fig \ref{fig:sketch}. This allows to exploit the advantage of graphene having a strong Coulomb interaction while the acoustic phonon coupling is weak in comparison to an ordinary two-dimensional electron gas in a semiconductor heterostructure. As a result, graphene under Landau quantization presents optimal conditions for the appearance of a significant carrier multiplication.

\textbf{Theoretical approach.} While the carrier dynamics in graphene has been intensively investigated over the last few years \cite{Dawlaty2008,SunNorris2008,Winnerl2011,Breusing2011,Johannsen2013}, there has been only a single study in the presence of a strong magnetic field by Plochocka et al. \cite{Plochocka2009}. In the latter study, the relaxation dynamics between higher Landau levels ($n\approx100$) is investigated in a differential transmission experiment revealing the importance of Auger scattering in a magnetic field and an overall suppression of scattering rates in comparison to the no-field case. Based on the density matrix formalism \cite{KochBuch,MalicBuch}, we perform time-resolved microscopic calculations of the ultrafast carrier relaxation dynamics within the energetically lowest Landau levels. We exploit a correlation expansion in second order Born-Markov approximation \cite{KochBuch,MalicBuch} to derive Bloch equations for graphene under Landau quantization: 
\begin{widetext}
\begin{eqnarray}
\dot{\rho}_{n}(t) & = & -2\sum_{n'}\text{Re}[\Omega_{nn'}(t)\, p_{nn'}(t)]+S_{n}^{\text{in}}(t)\left[1-\rho_{n}(t)\right]-S_{n}^{\text{out}}(t)\rho_{n}(t),\label{eq:Bloch-1}\\
\dot{p}_{nn'}(t) & = & i\triangle\omega_{nn'}p_{nn'}(t)+\Omega_{nn'}(t)\,\left[\rho_{n}(t)-\rho_{n'}(t)\right]-\frac{\Gamma(t)}{\hbar}p_{nn'}(t).\label{eq:Bloch-2}
\end{eqnarray}
\end{widetext}
It is a coupled system of differential equations for the carrier occupation probability $\rho_{n}(t)$ in the Landau level $n$ and the microscopic polarization $p_{nn'}(t)$ being a measure for optical inter-Landau level transitions according to optical selection rules, cf. Fig. \ref{fig:sketch}. The equations explicitly include the carrier-light interaction, all energy conserving electron-electron scattering processes, as well as the coupling with most relevant optical phonons. The carrier-light interaction enters the equations via the Rabi frequency $\Omega_{nn'}(t)=\frac{e_{0}}{m_{0}}\mathbf{M}_{nn'}\cdot\mathbf{A}(t)$ with the elementary charge $e_{0}$, the electron mass $m_{0}$, the optical matrix element $\mathbf{M}_{nn'}$, and the vector potential $\mathbf{A}(t)$. The energy difference $\triangle\omega_{nn'}=(\epsilon_{n}-\epsilon_{n'})/\hbar$ between the involved Landau levels defines the resonance condition. The magnetic field is incorporated into the equations by exploiting the Peierls substitution (
see Supplementary Note 1), accounting for the change of electron momentum induced by the confinement into cyclotron orbits \cite{Goerbig2011Review,Sipe2012}. The appearing in- and out-scattering rates $S_{n}^{\text{in/out}}$ describe Coulomb- and phonon-induced many-particle scattering processes. The rates include Pauli-blocking terms and the corresponding matrix elements, cf. Supplementary Notes 2-4 for more details. Due to the energy mismatch of the optical phonon modes with the inter-Landau level transitions at the assumed magnetic field strength, optical phonon scattering is strongly suppressed. Processes involving acoustic phonons and two-phonon relaxation processes can occur \cite{DasSarma2011Review,Malic2011,Li2013}, however, both are much slower compared to the Coulomb-induced dynamics and are neglected in the following. The scattering processes do not only change the carrier occupation $\rho_{n}(t)$, they also contribute to the dephasing $\Gamma$ of the microscopic polarization $p_{nn'}(t)$. The 
main contribution of $\Gamma$ stems from electron-impurity scattering, which also induces a broadening of the Landau levels. This broadening is explicitly calculated in the self-consistent Born approximation following the approach of Ando \cite{Ando1974_I}, cf. Supplementary Note 5, where the electron-impurity scattering strength is chosen in agreement with previous theoretical studies \cite{Ando1998,Levitov2012}. The validity of the applied Markov approximation has been tested by performing calculations taking also into account non-Markovian effects. The latter reflect the quantum-mechanical energy-time uncertainty principle giving rise to a weak temporal oscillation of carrier occupations. However, the effect is small and does not qualitatively change the investigated inter-Landau level carrier dynamics, cf. Supplementary Note 6.

In order to account for momentum-dependent dynamical screening, we consider the dielectric function $\epsilon(q,\omega)$ in the random phase approximation following the approach of Refs. \cite{Goerbig2011Review,Roldan2010}. While the real part of the dielectric function closely resembles the static limit (cf. Ref. \cite{Goerbig2011Review,Roldan2010}), taking a finite energy transfer $\hbar\omega$ into account gives rise to a non-vanishing imaginary part, cf. Supplementary Note 3 for more details. Then, the Fourier transform of the Coulomb potential $V_{q}$ appearing in the Coulomb matrix element 
\begin{eqnarray}
V_{34}^{12} & = &
\sum_{\mathbf{q}}V_{\mathbf{q}}\Gamma_{13}(\mathbf{q})\Gamma_{24}
(-\mathbf{q}),\label{eq:Coulomb matrix element Fourier representation}\\
\Gamma_{if}(\mathbf{q}) & = &
\int\text{d}\mathbf{r}\,\Psi_{f}^{*}(\mathbf{r})e^{
i\mathbf{q}\mathbf{r}}\Psi_{i}(\mathbf{r}),
\end{eqnarray}
is substituted by the screened potential $V_{q}/\epsilon(q,\omega)$. Here, $q$ and $\hbar\omega$ are the momentum and energy transfers of the scattering events and $\Psi_{i}$ denotes the wave function of a state $i$. Since in the presence of a magnetic field, the momentum is not a well-defined quantity, the vanishing screening in the long-wavelength limit does not imply a divergence of the Coulomb interaction. Furthermore, the singularities for collinear scattering processes appearing in graphene in the absence of a magnetic field \cite{Brida2013,TomadinPolini2013} do not occur in Landau-quantized graphene due to the absence of a strictly defined momentum conservation in such a discrete system. Moreover, note that excitonic effects of the Coulomb interaction can be neglected, since the Landau level broadening is generally larger than the Coulomb-induced finite excitonic dispersion, cf. Supplementary Note 3.

\begin{figure}[!t]
\begin{centering}
\includegraphics[width=85mm]{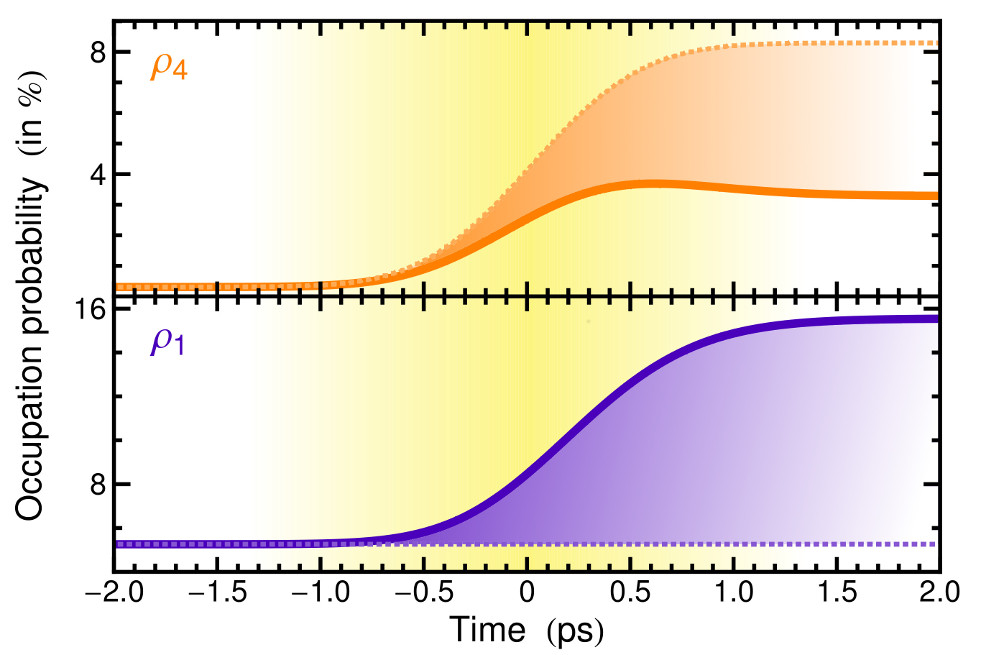} 
\par\end{centering}

\caption{\textbf{Landau level occupations.} Temporal evolution of the occupation probability $\rho_{4}(t)$ involved in the optical excitation characterized by an ultrashort pulse with a width of $\unit[1]{ps}$ and $\rho_{1}(t)$ reflecting the importance of Auger processes in the presence of a magnetic field with $B=\unit[4]{T}$. The dashed lines illustrate the dynamics in the absence of Coulomb scattering. The yellow shaded region denotes the width of the excitation pulse. The occupation $\rho_{2}(t)$ is not involved in the relaxation dynamics, $\rho_{0}(t)$ remains constant, and $\rho_{3}(t)$ shows an increase during pumping just like $\rho_{4}(t)$. The study is performed for undoped graphene at room temperature. }

\label{fig:occupations_0.1pump} 
\end{figure}

\textbf{Carrier relaxation dynamics.} Using tight-binding wave functions, we analytically determine the electronic bandstructure and the matrix elements appearing in the Bloch equations. Then, we have all ingredients at hand to evaluate the equations and obtain a microscopic access to the time-resolved ultrafast relaxation dynamics of optically excited carriers. The initial occupation probabilities $\rho_{n}(t=0)$ before the optical excitation are given by the Fermi function at room temperature. We apply an external magnetic field of $B=\unit[4]{T}$ and an optical excitation pulse with a width of $\unit[1]{ps}$, an applied pump fluence of $\epsilon_{\text{pf}}=\unit[10^{-2}]{\mu Jcm^{-2}}$, and an energy of $\hbar\omega_{L}\approx\unit[280]{meV}$ matching the inter-Landau level transitions $-3\,\rightarrow\,+4$ and $-4\,\rightarrow\,+3$, cf. Fig. \ref{fig:sketch}. As illustrated in Fig. \ref{fig:occupations_0.1pump}, the carrier occupation probability $\rho_{4}$ increases up to $8\%$ during the optical 
excitation (blue shaded area), while $\rho_{-3}$ decreases for the same amount (not shown). Interestingly, the build up of occupation in the Landau level $n=+4$ is stopped during the optical excitation and $\rho_{+4}$ starts to slightly decrease, cf. the orange solid line in Fig. \ref{fig:occupations_0.1pump}. This effect can be unambiguously traced back to the Coulomb-induced relaxation process, cf. the orange dashed line in Fig. \ref{fig:occupations_0.1pump} illustrating the temporal evolution of $\rho_{+4}$ without Coulomb scattering. At the same time, the occupation in Landau level $n=+1$ that is not driven at all by the optical pulse, increases until saturation is reached. The occupation $\rho_{1}$ remains constant, if the Coulomb interaction is switched off, cf. the blue dashed line in Fig. \ref{fig:occupations_0.1pump}. This behavior can be understood as following: First, the exciting laser pulse generates a non-equilibrium carrier distribution by transferring electrons from $n=-3$ to $n=+4$ (yellow 
arrows in Fig. \ref{fig:sketch}). Then, Auger-type processes including impact excitation (IE) and Auger recombination (AR) redistribute the carriers between the equidistant Landau levels $n=+4,\,+1,$ and $0$ (red and green arrows in Fig. \ref{fig:sketch}). Our calculations clearly reveal that IE is the predominant channel at the beginning of the relaxation dynamics due to the Pauli blocking terms in the scattering rates $S_{n}^{\text{in/out}}$ in the Bloch equations. During the optical excitation, the occupation of $n=+4$ is enhanced, whereas $n=+1$ contains only few electrons corresponding to the thermal distribution at room temperature. As a result, the electron-electron interaction favors the transition $+4\,\rightarrow\,+1$ over the inverse process and provides the energy to excite an additional electron into $n=+1$ from the energetically equidistant zeroth Landau level. This explains the observed increase of $\rho_{1}$ and the decrease of $\rho_{4}$, cf. Fig. \ref{fig:occupations_0.1pump}. Interestingly,
 the occupation $\rho_{0}$ does not change and remains $0.5$ during the entire relaxation dynamics (not shown). This is related to the peculiarity of the zeroth Landau level being half valence half conduction band \cite{Goerbig2011Review}. Due to the symmetry between electrons and holes in the considered undoped graphene, each transition evolving electrons can also occur with holes with exactly the same probability. As a result, the electron transition $0\,\rightarrow\,+1$ is always accompanied by a hole transition $0\,\rightarrow\,-1$ corresponding to the excitation of an electron into the zero Landau level. This results in an unconventional IE, where effectively only one charge carrier is created in every scattering event. The result is a constant $\rho_{0}$, while $\rho_{4}$ decreases and $\rho_{1}$ increases, just as our theory predicts.

\begin{figure}[!t]
\begin{centering}
\includegraphics[width=85mm]{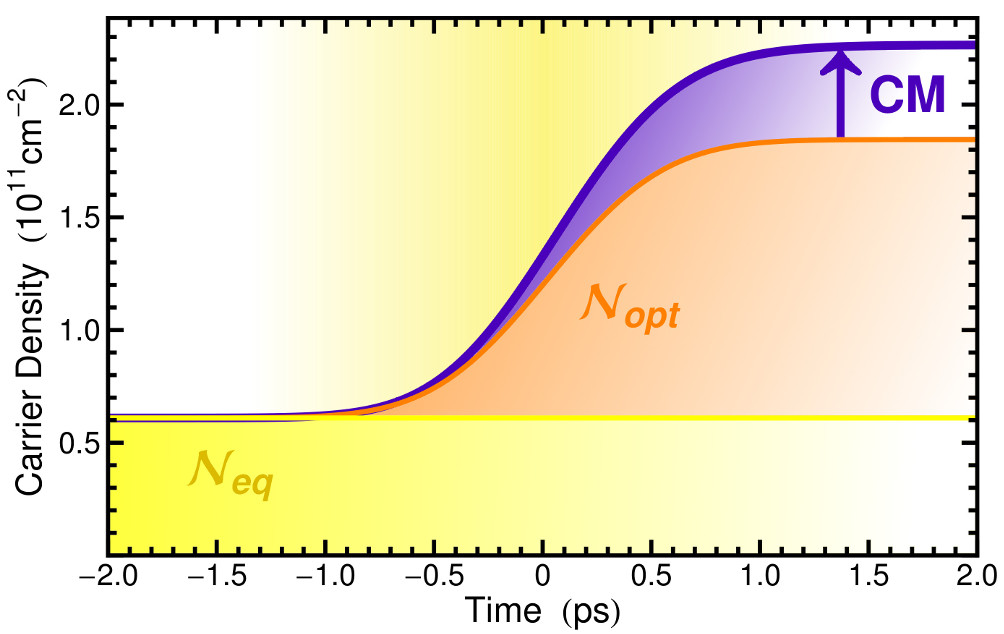} 
\par\end{centering}

\caption{\textbf{Carrier density.} Temporal evolution of the carrier density $\mathcal{N}$ including the thermal equilibrium distribution $\mathcal{N}_{\text{eq}}$, and the optically excited carrier density $\mathcal{N}_{\text{opt}}$. The Auger-induced carrier dynamics results in a carrier multiplication of $CM\approx1.3$ at the considered pump fluence of $\unit[10^{-2}]{\mu Jcm^{-2}}$.}

\label{fig:carrier density_0.1pump} 
\end{figure}

\textbf{Carrier multiplication.} To examine the total impact of Auger scattering processes, we investigate the temporal evolution of the carrier density that is composed of electrons in the conduction and holes in the valence band Landau levels. The electron-hole symmetry allows a convenient definition of the carrier density $\mathcal{N}=2(N_{\text{deg}}/A)\sum_{n\geq1}\rho_{n},$ with the factor 2 accounting for the electron-hole symmetry, the total degeneracy of each Landau level $N_{\text{deg}}$ and the area $A$. Under the investigated conditions, the zeroth Landau level, being equally divided between both bands, has a temporally constant occupation and does not contribute. Due to the finite temperature, the initial value of the carrier density is non-zero and we start with an equilibrium value $\mathcal{N}_{\text{eq}}$ (yellow line in Fig. \ref{fig:carrier density_0.1pump}). The optical pulse excites carriers and therewith leads to an ascent of the carrier density on a picosecond time scale. If no many-
particle interactions are taken into account, $\mathcal{N}_{\text{opt}}$ reaches a value of approximately $\unit[1.8\cdot10^{11}]{cm^{-2}}$, cf. the orange line in Fig. \ref{fig:carrier density_0.1pump}. The full dynamics including Coulomb-induced carrier scattering reveals a further increase of the carrier density up to $\unit[2.3\cdot10^{11}]{cm^{-2}}$ reflecting the appearance of a significant carrier multiplication. The CM factor is defined by 
\begin{equation}
CM=\frac{\mathcal{N}-\mathcal{N}_{\text{eq}}}{\mathcal{N}_{\text{opt}}},
\end{equation}
and corresponds to $1.3$ at the given characteristics of the excitation pulse. It is a direct consequence of the impact excitation, where more carriers are generated in $n=+1$ than are lost in $n=+4$. This is apparent when comparing the drop-off in $\rho_{4}$ ($5\%$) with the increase of $\rho_{1}$ ($10\%$) after two picoseconds in Fig.
\ref{fig:occupations_0.1pump}.

\begin{figure}[!t]
\begin{centering}
\includegraphics[width=85mm]{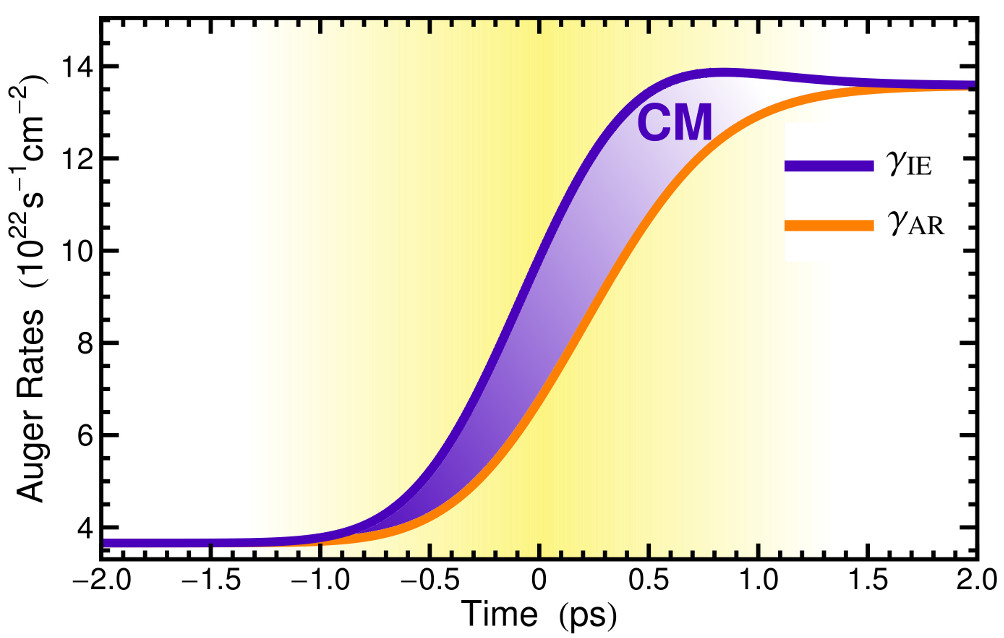} 
\par\end{centering}

\caption{\textbf{Coulomb scattering rates.} Scattering rates $\gamma_{\text{IE/AR}}$ for the impact excitation (IE) and Auger recombination (AR). The yellow area represents the excitation region. The blue shaded area illustrates the asymmetry region in which IE prevails over AR resulting in the occurrence of carrier multiplication (CM).}

\label{fig:Auger rates_0.1pump} 
\end{figure}

To further prove the occurrence of CM as direct implication of an efficient IE, we determine the time-resolved scattering rates $\gamma_{\text{AR/IE}}(t)=1/A\sum_{i}\dot{\rho}_{i}(t)\Big|_{\text{AR/IE}}$ of the two competing Auger channels. Our calculations show a much more efficient IE as a consequence of the non-equilibrium distribution created by the optical excitation, cf. Fig. \ref{fig:Auger rates_0.1pump}. The ongoing impact excitation during and after the excitation leads to a more efficient Auger recombination until both rates are equal after roughly $\unit[2]{ps}$. This is the time frame, in which CM takes place, cf. Fig. \ref{fig:carrier density_0.1pump}. Its value corresponds to the area between both curves, which is ultimately defined by the asymmetry between the two competing processes. Their efficiency as Coulomb channels strongly depends on the pump fluence determining the number of available scattering partners. Furthermore, the electronic temperature and the strength of the magnetic field 
have an influence on the predicted asymmetry between IE and AR.

\textbf{Discussion.} Addressing the optimization of CM, we investigate its
dependence on the externally controllable parameters, such as pump fluence,
temperature, and magnetic field, cf. Fig. \ref{fig:CM(fluence,B,T)}. Although
the initial asymmetry between impact excitation and Auger recombination is
larger at enhanced pump fluences, it also leads to a faster equilibration of
both rates leaving a shorter time frame for CM to occur \cite{Winzer2012}. As a
result, the carrier multiplication decreases with the pump fluence until a
saturation of pumping at higher fluences is reached, as illustrated in Fig.
\ref{fig:CM(fluence,B,T)}a. Then, CM starts to slightly increase again
reflecting the inefficient optical pumping at saturation conditions. The
dependence on temperature is governed by occupation effects. A higher
temperature results in less efficient impact excitation, since an increased
initial occupation of $n=+1$ lowers the efficiency of the electronic transition
$0\,\rightarrow\,+1$. The qualitative temperature trend 
resembles a Fermi distribution (cf. Fig. \ref{fig:CM(fluence,B,T)}b) resulting in the largest CM for temperatures lower than $200K$. Finally, the dependence on the magnetic field is more complex as it is a combination of both effects discussed above, namely the excitation strength and occupation effects. Increasing the magnetic field, the Landau levels (except for $n=0$) shift to higher energies according to the relation $E_{n}\propto\pm\sqrt{B\left|n\right|}$. Therefore, the initial thermal occupation probability $\rho_{n}$ for $n=1,2,3...$ is reduced with increasing magnetic field and IE becomes more efficient. Moreover, the effective excitation strength is determined by the amplitude of the laser pulse and is inverse proportional to its frequency.
\begin{figure}[!t]
\begin{centering}
\includegraphics[width=85mm]{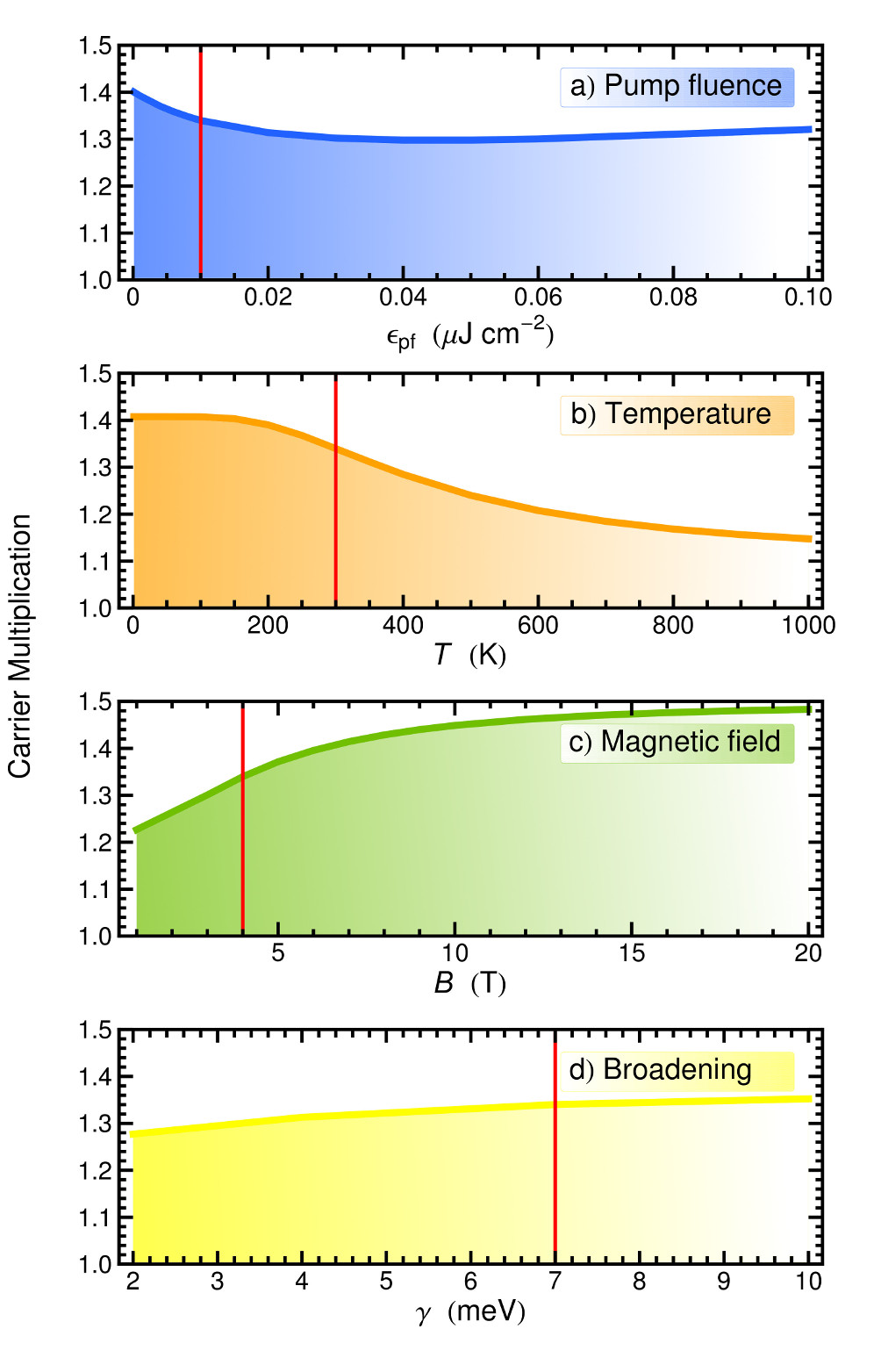} 
\par\end{centering}

\caption{\textbf{Optimization of carrier multiplication.} Carrier multiplication dependence on (a) pump fluence, (b) temperature, (c) magnetic field, and (d) Landau level broadening. The red lines represent the values used in Figs. \ref{fig:occupations_0.1pump}-\ref{fig:Auger rates_0.1pump}.}

\label{fig:CM(fluence,B,T)} 
\end{figure}
 As the latter increases with the magnetic field, the pumping becomes less efficient giving rise to a more pronounced CM, as already shown in Fig. \ref{fig:CM(fluence,B,T)}a. We observe a saturation value for the carrier multiplication of $1.5$, cf. Fig. \ref{fig:CM(fluence,B,T)}c. This value is characteristic for the investigated undoped graphene within the energetically lowest Landau levels. Here, IE can only occur once via scattering involving the energetically equidistant Landau levels $n=+4,\,+1,$ and $0$. Furthermore, the peculiar nature of the zeroth Landau level being a conduction and valence band at the same time allows only the generation of one additional charge carrier per excitation, as discussed above. Therefore, an optically excited electron-hole pair can yield in total up to three charge carriers resulting in a maximal CM of $1.5$. Note that at higher magnetic fields, certain inter-Landau level transitions can become resonant with optical phonon modes creating so called magneto-phonon 
resonances. This is expected to reduce the observed CM, since the phonons open up a new relaxation channel directly competing with Auger processes, and even the electron-phonon interaction is modified \cite{Ando2007,Goerbig2007,Faugeras2009,Yan2010}. Therefore, the high-magnetic field region of Fig. \ref{fig:CM(fluence,B,T)}c should be understood as a trend of the CM as a function of the magnetic field. At very low magnetic fields, the spacing between Landau levels decreases, resulting in an overlapping of levels. Here, the applied theoretical approach is not valid anymore. Furthermore, note that the carrier multiplication does not significantly depend on the Landau level broadening, as illustrated in Fig. \ref{fig:CM(fluence,B,T)}d. On the one hand, the Lorentzian appearing in the scattering rates decreases for larger broadenings. On the other hand, the dielectric function is reduced leading to an enhancement of the Coulomb interaction. Additionally, due to an increased dephasing of $p_{nn'}$, the optical 
pumping is less efficient. Overall these effects nearly cancel out resulting in a CM that is to a large extent independent of Landau level broadening.

Including energetically higher Landau levels will increase the CM, since the excess energy of a highly excited electron-hole pair can provide enough energy for more than one IE scattering event. In this very first study, we have provided clear evidence of CM in the low energy limit. Taking into account the findings of Plochocka et al. \cite{Plochocka2009} that experimentally prove the importance of Auger scattering even for higher Landau levels, we expect a pronounced CM to occur in a broad spectral range. Although they suspect that Auger scattering may be suppressed for lower Landau levels, their findings are indeed compatible with our results. A multi-color pump-probe experiment addressing the carrier dynamics between the discrete low-energy Landau levels should confirm our theoretical predictions in a straightforward way. 

In summary, we have gained new fundamental insights into the relaxation dynamics in graphene under Landau quantization, in particular revealing the appearance of a considerable carrier multiplication that can be traced back to an efficient impact excitation. We have shown that Landau quantization offers a promising strategy to address the key challenge of carrier extraction in a gapless structure which might be beneficial for the design of graphene-based optoelectronic devices.

\section*{Acknowledgments}

We acknowledge the financial support from the Einstein Stiftung Berlin
and we are thankful to the DFG for support through SPP 1459. Furthermore,
we thank M. Helm, S. Winnerl, M. Mittendorff (Helmholtz-Zentrum Dresden-Rossendorf),
T. Winzer, G. Berghaeuser, and Eike Verdenhalven (TU Berlin) for fruitful
discussions. A.K. thanks Sfb 787 and the BMBF (Nanopin) for support.

\end{document}